\documentclass[aps,showpacs,pra,twocolumn,superscriptaddress]{revtex4}
\usepackage{graphicx}
\usepackage{color}
\usepackage{dcolumn}
\usepackage{bm}
\usepackage{graphicx}
\usepackage{amsmath}

\usepackage[USenglish]{babel}

\begin{document}
\title{Tuneable defect interactions and supersolidity in dipolar quantum gases on a lattice potential}

\author{Wolfgang Lechner}
\email{w.lechner@uibk.ac.at}
\affiliation{Institute for Quantum Optics and Quantum Information, Austrian Academy of
Sciences, 6020 Innsbruck, Austria}
\affiliation{Institute for Theoretical Physics, University of Innsbruck, 6020 Innsbruck,
Austria}
\author{Fabio Cinti}
\email{cinti@sun.ac.za}
\affiliation{National Institute for Theoretical Physics (NITheP), Stellenbosch, South Africa}
\author{Guido Pupillo}
\email{pupillo@unistra.fr}
\affiliation{icFRC, IPCMS (UMR 7504), ISIS (UMR 7006), Universit\'e de Strasbourg and CNRS, 67000 Strasbourg, France}

\date{\today }


\begin{abstract} 
Point defects in self-assembled crystals, such as vacancies and interstitials, attract each other and form stable clusters. This leads to a phase separation between perfect crystalline structures and defect conglomerates at low temperatures. We propose a method that allows one to tune the effective interactions between point defects from attractive to repulsive by means of external periodic fields. In the quantum regime, this allows one to engineer strongly-correlated many-body phases. We exemplify the microscopic mechanism by considering dipolar quantum gases of ground state polar molecules and weakly bound molecules of strongly magnetic atoms trapped in a weak optical lattice in a two-dimensional configuration. By tuning the lattice depth, defect interactions turn repulsive, which allows us to deterministically design a novel supersolid phase in the continuum limit. 
\end{abstract}

\maketitle
\section{Introduction}

Defects are crucial for the determination of macroscopic mechanical, optical, and electronic properties of solids \cite{TAYLOR,DISLOCATIONS,HIRTH_LOTHE}. One key aspect is the mutual effective interactions between point defects such as vacancies or interstitials, corresponding to the lack or excess of crystal particles, respectively. In self-assembled classical crystals the effective interactions between point defects is attractive for all combinations of defects (vacancy-vacancy, interstitial-interstitial, interstitial-vacancy) in a wide range \cite{CLASSICALDEFECT2,DEFECTS_CLASSICAL,L_DISPLACEMENT,L_ELASTCITY}. This interaction leads to the formation of string-like defect clusters \cite{COLLOIDEXP}. The mechanism behind the attraction of defects is a result of non-linear effects in the displacement fields of multiple defects and cannot be described as a simple pair interaction within elasticity theory. In the quantum regime, the interaction between vacancies is even less well understood, and it is an open question whether the classical results may be directly used to infer many-body properties in the {\it quantum} regime. There, the interaction between vacancies is a crucial part of the theory for the supersolid phase \cite{ALC,ALC2,GROSS,ENIGMA,ANDERSON,BoninsegniRevModPhys}. The supersolid is conjectured to be a result of delocalized vacancies. However, the precise role of the defect dynamics to establish both superfluid and crystalline orders has been the object of intense investigations in the last decades \cite{SUSOEXP,DEFECTS_FATE,Boninsegni2005,KIMCHAN,CINTI,SOFTCORE2,SOFTCORE1,REATTOREV,RICA1,RICA2,MACRI1,MACRI2,KUNIMI,ANCILOTTO}. 

\begin{figure}[t]
\centerline{\includegraphics[width=6cm]{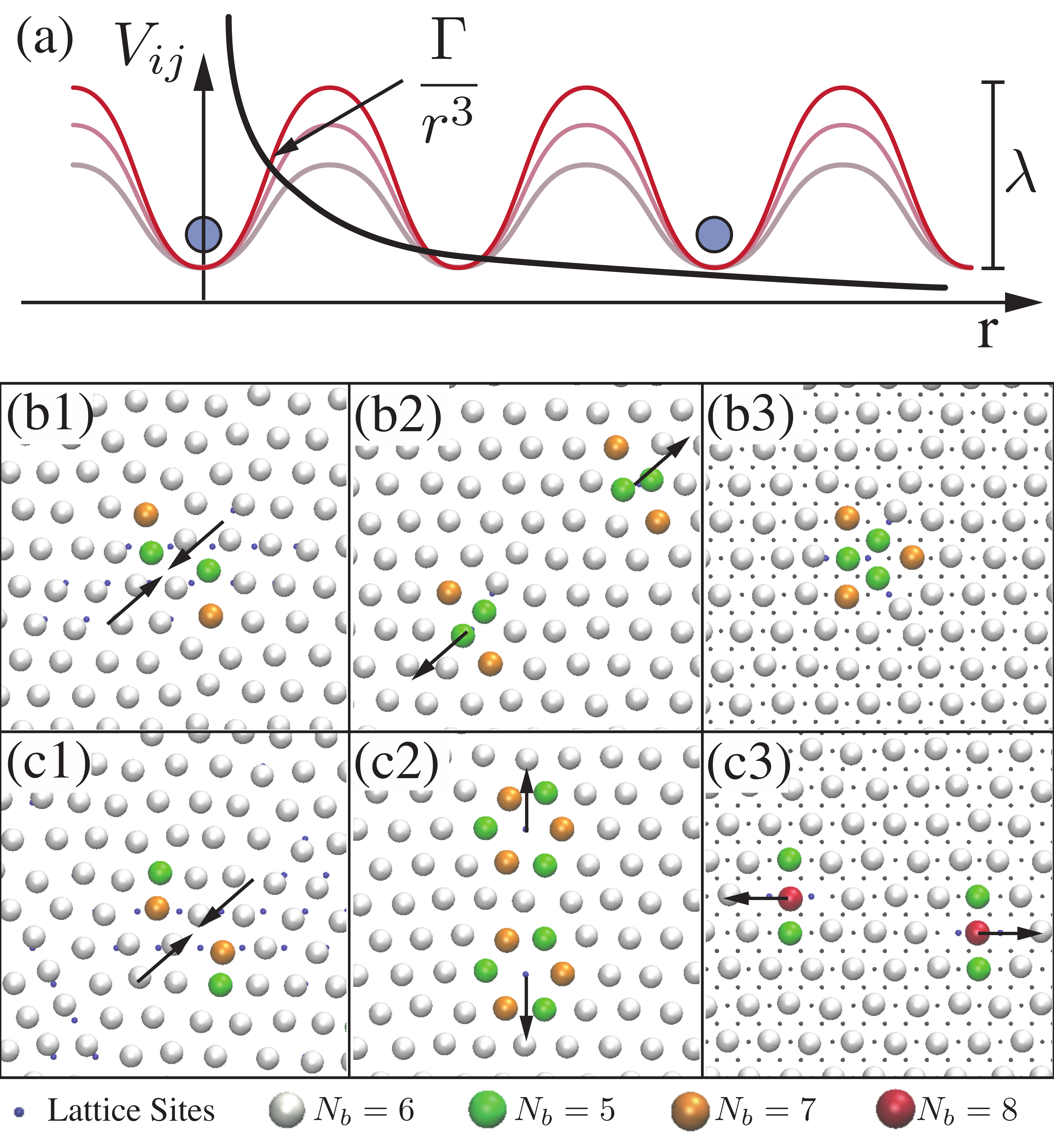}}
\caption{(a) Sketch of setup: Particles are prepared in a two-dimensional self assembled crystal with an additional quasi-commensurate optical lattice with varying strength $\lambda$. Particles are colored by their numbers of nearest neighbors $N_b$. {\it Interstitials}: (b1) Classical steady state of two interstitials for $\lambda=0$. In the absence of an optical lattice, defects collapse to dislocation pairs which are attractive (arrows) and form a stable pair in a defect string \cite{L_INTERACTION}. Dots indicate the position of the undistorted lattice; (b2) For a deep lattice $\lambda=6$ with spacing $a_0$ the interaction between interstitials is purely repulsive; (b3) Interstitials for lattice spacing $a_0/2$ and $\lambda = 6$ can form a bound pair with triangular order. {\it Vacancies}: (c1) Vacancies in a self-assembled crystal and $\lambda=0$ form defect strings, similar to interstitials \cite{L_INTERACTION}; (c2) Vacancies in a deep lattice with $\lambda=6$ and spacing $a_0$ (c2) and $a_0/2$ (c3) are purely repulsive.}
\label{fig:fig1}
\end{figure}

Here, we show that effective interactions between point defects in a self-assembled crystal can be systematically tuned from attractive to repulsive by means of external periodic superlattices. The lattice spacing of the optical lattice is chosen to be identical or double the lattice spacing of the self assembled crystal. In this setup, the interaction between point defects is the 
result of an interplay between displacement induced attraction and energy induced repulsion. By properly choosing the parameters of the superlattice, the relative strength of the displacement-induced energy and entropy parts can be tuned, thus making the interaction an experimentally accessible variable.  

Based on this microscopic picture of tunable defects, we investigate the many-body quantum regime. As an example, using exact quantum Monte-Carlo methods, we study the phases of a dipolar bosonic gas in the presence of a weak triangular optical lattice. Unlike previous works that focused on the tight-binding limit valid for deep lattice potentials~\cite{Danshita,Pollet2010,Boninsegni2005,BoninsegniRevModPhys}, here we focus on the {\it continuous space limit} where the band structure of the lattice is not formed or barely formed. We investigate the phase diagram by varying the lattice depth and the strength of the dipole-dipole interaction both for a commensurate and an incommensurate filling of the lattice potential, around a small density $n=1/4$. 
In the commensurate case, we observe a superfluid to insulating quantum phase transition. 
Most notably, introducing defects in the presence of the weak periodic potential allows for the realisation of defect-induced supersolidity as originally proposed by Andreev and Lifshitz~\cite{ALC2}. In our case, the triangular crystalline structure for the dipoles is either due to direct strong dipolar interactions or to "pinning" of the weakly interacting superfluid by the lattice potential (without opening of a spectral gap), depending on the system parameters. We present a zero-temperature phase diagram which furnishes a complete description of the superfluid-supersolid and crystal-supersolid phase transition.
We find that superfluity and $quasi$-condensate fraction are comparatively robust against finite temperature, and estimate possible experimental parameters for experiments with ground state polar molecules~\cite{recentJILA,JILA,MOL1,MOL2,MOL3,LICS,NAK,HEXTRANS} and weakly bound molecules of strongly dipolar magnetic atoms~\cite{Ferlaino,Laburthe-Tolra,Pfau,Lev}. These results differ qualitatively from their deep-lattice counterparts, where a superfluid state was found for these low densities~\cite{Danshita,Pollet2010}.

We note that investigations of the quantum mechanical phases of lattice Hamiltonians in the continuum limit have so far focused on {\it contact} interactions \cite{Buechler, Astra2007,Filinov}. This has led to the prediction of generalized superfluid-Mott insulator transitions in continuous space \cite{Pilati}, as well as of solid and superfluid phases of, e.g, He, He$_2$ and D$_2$ adsorbed on solid-state surfaces such as, e.g., graphite \cite{Pollet}, graphane \cite{Nava,Carb,Boronat2009,Boronat2010}, Alkali substrates \cite{ALKALISUBSTRATE,Caz2013}.  The present work takes a step in a similar direction for the case of {\it finite range} interactions.\\

The methodology presented in this paper introduces a new toolbox for the manipulation of complex matter, in analogy to the techniques developed  to tune and shape the direct interactions between particles in systems as diverse as classical colloids \cite{Likos2001,Int0,Int1,Int2,Int3,Int4,Int5,BECHINGER}, as well as atomic and molecular systems in the quantum regime \cite{RYD1,Chin2010,Jones2006,Buechler,Gorshkov2013,RYD3}, which is the basis for the success in the realization of many-body phases in these systems~\cite{BlochRMP,Baranov,Tura}. The proposal is based on self-assembled 2D crystals of polar molecules. For details on the required parameters for trapping and interaction strengths see Ref. \cite{Buechler}.  \\

The paper is organised as follow: in  Section~\ref{Model} we introduce the model Hamiltonian. In Section III we show and discuss 
in detail results concerning defect interactions in the classical regime. The study of the quantum phases is proposed in Section IV.
Section V discusses the influence of temperature on supersolidity as well as parameters for possible physical realisations of self-assembled crystals in cold dipolar quantum gases \cite{Buechler}. 
Finally, in Section VI we draw some conclusions and future outlooks. 

\section{Model}\label{Model}

We consider ultracold polar molecules which are trapped in two dimensions with an additional triangular optical lattice [Fig.~\ref{fig:fig1}(a)] described by the Hamiltonian 
\begin{equation}
\label{hamiltonian}
H = \frac{1}{2m}\sum_i {\bf p}_i^2 + \sum_{i<j} V_{ij} + \lambda \sum_i U_i(x,y).
\end{equation}
Here $m$ is the mass, while $\mathbf{p}_i$ represents the single particle momentum. 
In the presence of a perpendicular electric field, the dipole-dipole interaction 
$V_{ij}  = \Gamma_0/r^3$ is purely repulsive 
\cite{Buechler} with strength $\Gamma_0$. 
The last term of Eq.~(\ref{hamiltonian}) represents the external potential of a triangular optical lattice with depth $\lambda$ 
\begin{eqnarray}
\label{lattice}
U(x,y) =  &-&\sin \left[\pi (x+\sqrt{3} y)/ a_0 \right]^2 \\ \nonumber 
  &-& \sin \left[ \pi (- x + \sqrt{3} y)/ a_0 \right]^2 -  \sin \left[2 \pi x/a_0 \right]^2,
\end{eqnarray}
which can be implemented with two standing laser beams \cite{SUPERLATTICE}. 
In the following energies and distances are given in units of $\Gamma \equiv \Gamma_0/a_0^3$, 
$a_0$ being the optical lattice's constant. 
The particle density $\rho$ is given in units of $N/(A a_0^2 \sqrt{3}/2)$, where $A$ is the size of the system and $N$ the particle number. 
As a consequence, the filling fraction is $n=1$ for particles with average distance $a = a_0$ and $n=1/4$ for $a = 2 a_0$, respectively. We consider a crystal commensurate with respect to the lattice if the ratio between density and filling fraction is an integer value and the crystal has triangular symmetry. Therefore, both $n=1$ and $n=1/4$ are commensurate configurations. 

Hamiltonian Eq.~(\ref{hamiltonian}) can be realized with cold polar molecules (e.g. KRb~\cite{JILA}, RbCs~\cite{Innsbruck}, NaK~\cite{NAK} or LiCs~\cite{LICS}), where particle densities are usually $\rho \ll 1$~\cite{Ye2013}. Alternatively, Rydberg-dressed atoms \cite{Biedermann2014,Pfau2014,RYD1,RYD2,RYD3,Pohl2010,Pohl2014} or ground state magnetic atoms~\cite{Ferlaino,Laburthe-Tolra,Pfau,Lev} may be used, with $\rho\simeq1$. In the latter case, recent experiments with weakly bound Er$_2$ molecules composed of two highly-magnetic ground state atoms open the way to the realisation of cold quantum gases with comparatively large dipole moments and high densities $\rho\simeq1$~\cite{Dalmonte,Ferlaino2015}. 

\begin{figure}[t]
\centerline{\includegraphics[width=8.cm]{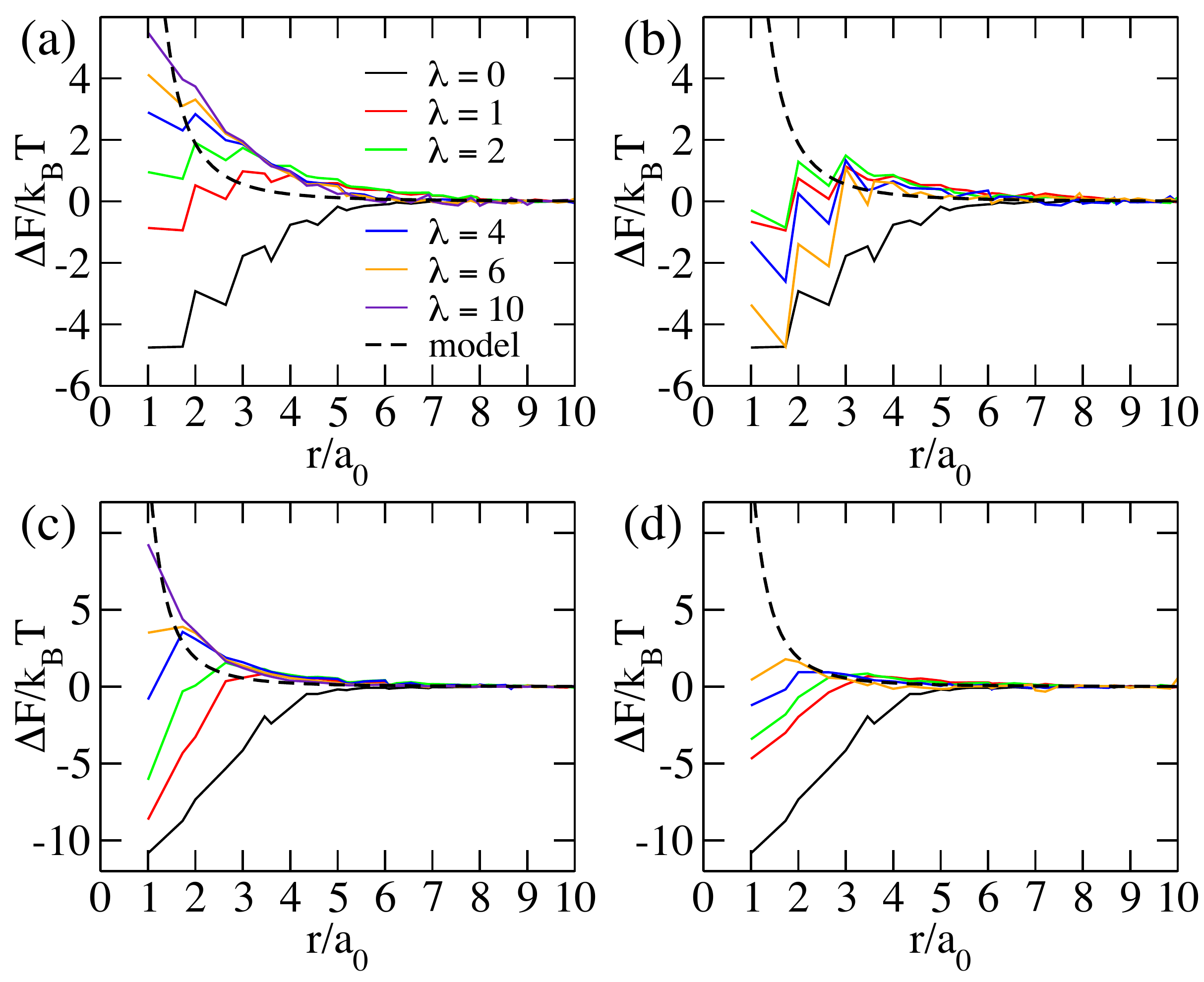}}
\caption{Effective interaction $V_{\rm eff}$ between interstitials and vacancies in a dipolar crystal as a function of the distance $r$ for various $\lambda$ and $\Gamma=15$ from a classical Monte Carlo simulation.  (a)  For $\lambda=0$, $V_{\rm eff}$ for interstitials is purely attractive (black). With increasing $\lambda$ and lattice filling $n=1$, the interaction can be tuned from attractive to purely repulsive. The dashed line is the analytical prediction $\Delta F/k_B T = \Gamma/r^3$ for a lattice model. (b) $V_{\rm eff}$ between interstitials can also be tuned with $\lambda$ in a lattice with filling $n=1/4$ (which corresponds to a lattice spacing $a_0/2$). For $\lambda \lesssim 6$ the shape of the interaction is similar to the filling $n=1$. For $\lambda \simeq 6$, however, a bound defect configuration becomes meta-stable [compare with Fig. \ref{fig:fig1}(b3)]. (c) Tunability of vacancy interactions for lattice filling $n=1$.  $V_{\rm eff}$ changes sign with increasing $\lambda$. (d) Tunability of vacancy interactions for lattice filling close to $n=1/4$. Note that for vacancies and $\lambda > 6$ hopping is basically suppressed. For interstitials, due to the pair interaction a sampling up to $\lambda = 10$ is possible. Note, that the roughness of the interaction is an artefact of the projection of a two-dimensional interaction to 1D. The two-dimensional potential surface for $\lambda=0$ is shown in Ref. \cite{L_INTERACTION}.}
\label{fig:results}
\end{figure}

\section{Defect Interactions}

Interstitials (vacancies) are point defects which result from adding (removing) a single particle to (from) a self-assembled crystal. Due to entropy, a finite number of defects  exists even in thermal equilibrium \cite{HIRTH_LOTHE}. These point defects induce a long-range displacement field \cite{L_DISPLACEMENT,L_ELASTCITY} with a non-linear short range part that is responsible for the complex dynamics of isolated defects \cite{REICHHARDT} and the interaction between them \cite{DEFECTS_CLASSICAL,L_INTERACTION}. On a quantitative level, effective interactions can be understood as the change in free energy, $\Delta F$, as a function of the defect distance $|{\bf r}|\equiv r$ for various $\lambda$ and fixed parameters ($\Gamma$, $n$, $a_0$) with
\begin{equation}
\label{eq:freeenergy}
V_{\text{eff}}(r) = \Delta F(r)/(k_B T)  = -\ln \langle \delta(r[\mathbf{x}] - r)\rangle.
\end{equation}
Here, $\langle \cdot \rangle$ denotes ensemble averages and $P(r)= \langle \delta(r[\mathbf{x}] - r)\rangle$ is the probability to find two defects at distance $r$. The first equality corresponds to the so-called \textit{reversible work theorem} \cite{CHANDLER}. 

The parameter $r$ in Eq.~\eqref{eq:freeenergy} describes the distance between the two defects and is determined using the following protocol, already introduced e.g. in Refs. \cite{L_INTERACTION,COLLOIDEXP}. In each time-step, a virtual triangular lattice with lattice spacing equivalent to the average particle distance is considered. Then, each particle is associated with the closest virtual lattice site. This implies that in the presence of an interstitial there will be exactly one doubly-occupied virtual lattice site. The position of this lattice site is defined as the position of the interstitial. For vacancies, the unoccupied lattice site is the position of the vacancy. The free energy profile of two defects as a function of the distance $r$ corresponds to the effective interaction between the two defects. In the following, Eq.~(\ref{eq:freeenergy}) is evaluated using Monte Carlo sampling in combination with the self-consistent histogram method (see Ref.~\cite{FRENKEL}). The effective force between the defects is then the negative slope of the free energy. 

In free space ($\lambda=0$) vacancies and interstitials attract each other in all combinations (vacancy-vacancy, interstitial-interstitial) and form string-like defect clusters, as shown in Ref. \cite{COLLOIDEXP}. Examples of this behaviour are given in  Figs.~\ref{fig:fig1} (b1) and (c1) for two interstitials and two vacancies, respectively. The resulting effective interaction potential $V_{\rm eff}$ is purely attractive and increases monotonically with $r$ for all cases, as shown in Fig. \ref{fig:results}(a-d) for $\lambda=0$ (black continuous line).
 
We find that the presence of an additional optical lattice [Fig.~\ref{fig:fig1}(a)]  changes dramatically the energetics, dynamics and interaction of defects. In particular, by increasing the lattice depth $V_{\rm eff}$ can become repulsive.  Example results for $V_{\rm eff}(r)$ between interstitials and vacancies for various lattice depths are shown in the upper and lower panels of Fig.~\ref{fig:results}, respectively, for two choices of particle densities $n=1$ [panels (a) and (c)] and  $n=1/4$ [panels (b) and (d)].  In all cases, the figure shows that for interstitials the turning point where  $V_{\rm eff}$ turns first from attractive to repulsive is $\lambda \approx 1$, while for vacancies a larger depth of $\lambda \approx 4$ is required. For comparatively large lattice depths (e.g., $\lambda > 6$) the interaction starts to approach the black dashed lines, which correspond to analytical results from a discrete lattice model introduced below. The sign and strength of the effective interactions is however density dependent for intermediate lattice strengths. For example, in the case of interstitials $V_{\rm eff}$ displays a non-monotonic dependence on $\lambda$ for densities close to $n=1/4$  [panel(b)]. For the dynamics of defects, this implies phase separation and cluster formation for $2\lesssim \lambda \lesssim 6$. An example of this behaviour is given in Fig.~\ref{fig:fig1} (b3). Finally, we find that for $\lambda\rightarrow \infty$ the dynamics of defects is effectively frozen, as the energy necessary to hop from one site to the next becomes increasingly prohibitive.

The limit of a large lattice depth in the commensurate crystal (with any filling) can be understood from the following simple lattice model. In this model, particles are fixed to lattice positions and can hop between sites with given (temperature-dependent) rates. Each lattice site can be un-occupied, occupied or doubly-occupied. As above, we assume that the direct interaction between individual particles is $V_{ij} = \Gamma/r^3$. We remove the divergence at $r=0$ by fixing the energy of a doubly-occupied site to $U$.

Let us first consider the interaction between two interstitials in this model. In this case, all sites will be occupied and two sites are doubly-occupied, for the case of unit filling $n=1$ plus two additional particles. The energy of the system is then $E_{\rm II} = 1/2 \sum_{i}^{N+2} \sum_{j<i} V_{ij} + 2 U$. The first term corresponds to the sum of all interactions in the system with $N+2$ particles and the second term is the $2U$ offset from the two interstitials. The first term is a function of the distance between the additional particles, say $N+1$ and $N+2$. The effective potential between the defects reads $V_{\rm eff}(r) = E_{\rm II}(r \rightarrow \infty) - E_{\rm II}(r)$. The $2U$ term and all contributions in the first term up to last term $V_{N+1,N+2}$ cancel. This is identical to the interaction between two particles and therefore, the effective potential is $V_{N+1,N+2} =  V_{\rm eff} = \Gamma/r^3$. 

For vacancies, the situation is less obvious. In this case, the system contains $N-2$ occupied lattice sites and two un-occupied lattice sites. Here, the distance between the two vacancies is defined as the distance between the two un-occupied lattice sites. Considering again $V_{\rm eff}(r) = E_{\rm VV}(r\rightarrow\infty) - E_{\rm VV}(r)$, we find that when summing up all energy contributions in the system the energy $E_{\rm VV}(r)$ behaves exactly as $V_{\rm eff}(r) = V_{ij}(r)$. Therefore, also two vacancies in the lattice are repulsive with $\Gamma/r^3$. These exact results are depicted in Fig. \ref{fig:results} as dashed lines. \\

Based on this microscopic classical model of tunable defect interactions, in the next Section we present results on defect-induced quantum phases using exact quantum Monte Carlo simulations for bosonic dipoles.

\begin{figure}[t]
\centerline{\includegraphics[width=0.9\columnwidth]{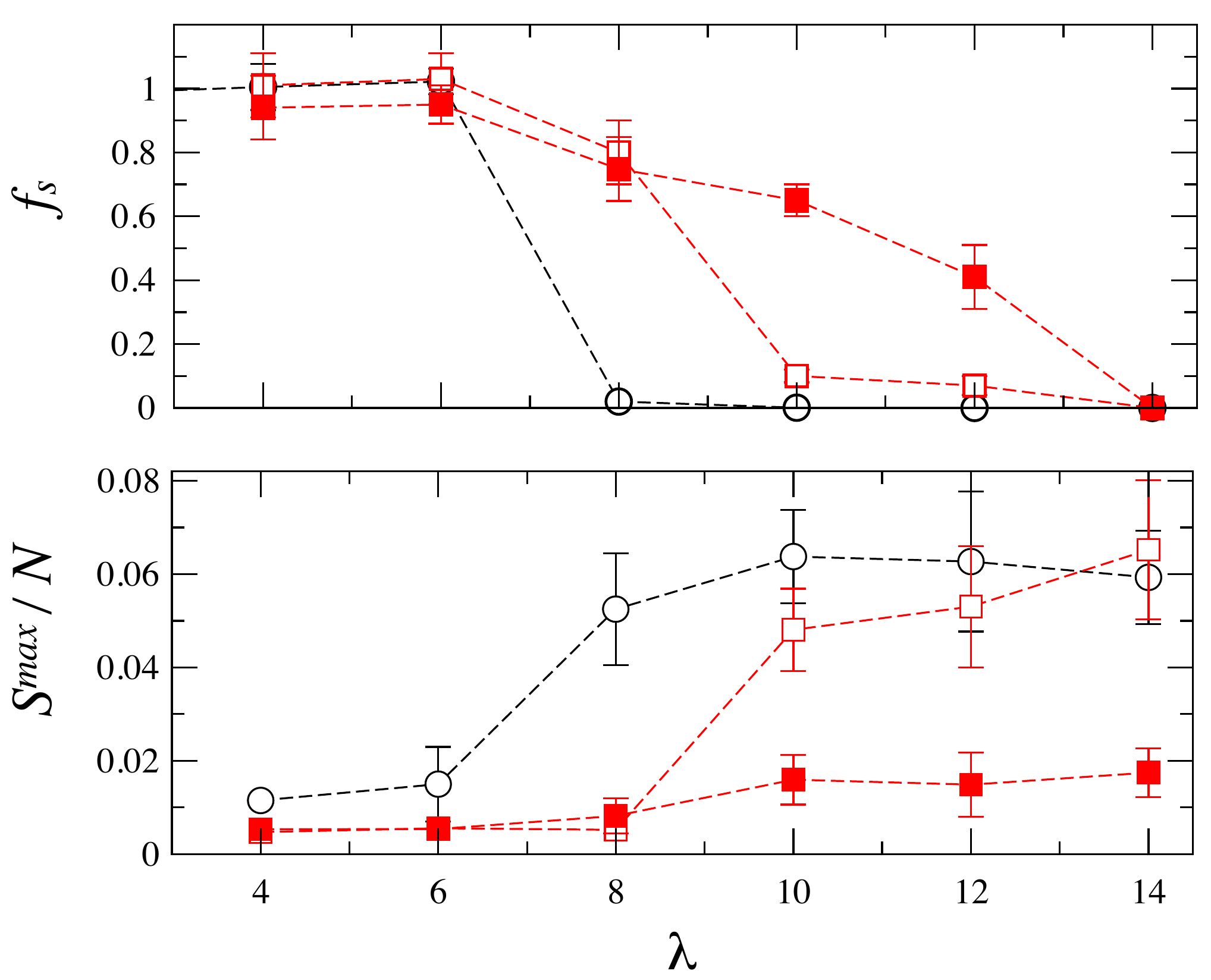}}
\caption{(Colour online) Superfluid fraction (top panel) and Structure factor (lower panel) as a function of the lattice depth $\lambda$ 
with $r_d\approx7.5$ for filling $n$=1/4, i.e. without defects ($N_d=0$, circle), and for $n\gtrsim1/4$, i.e. with 
defects: $N_d=4$ (full square, interstitial density 0.02), and $N_d=8$ (open square, interstitial density 0.04).}
\label{fig:sff1}
\end{figure}
 
\section{Quantum Phases}

The present section examines the applicability of classical predictions to the quantum regime,
considering commensurate and incommensurate filling values around $n=1/4$. 
We chose this value as the classical groundstate configuration at $n=1/4$ 
is the commensurate triangular crystal that best minimises lattice-induced frustration: the free-space triangular lattice is very little distorted by the external lattice potential. In addition, previous studies using tight-binding models for bosonic particles valid in the limit of very large lattice depths have mainly focused on experimentally challenging (for molecules) higher densities as $n >1/3$ \cite{Pollet2010, Boninsegni2005, Wessel2005, Heid2005, Melko2005}. This regime of lower densities is thus essentially unexplored from the point of view of the investigation of quantum many-body phases. 
 In the following we will be specifically concerned with the emergence of supersolid behaviour for densities close to the commensurate filling  $n=1/4$.\\

In our analysis, we use an exact numerical quantum Monte-Carlo algorithm in the continuous-space path integral (PIMC) representation \cite{WORM,Ceperley,Buechler}. Our PIMC code is based on the so-called worm algorithm, which is known to efficiently provide numerically exact estimates of thermodynamic quantities such as the superfluid density and the structure factor, which can be used as order parameters to determine the nature of superfluid and solid phase, respectively. The superfluid density reads
\begin{equation}
f_s= \frac{mk_B T}{\hbar^2 N}\langle {{\bf w}}^2 \rangle,
\end{equation}
with ${\bf w}=(w_x,w_y)$ the winding number estimator along the $x$ and $y$ directions \cite{SF}. 
The static structure factor is instead defined as 
\begin{equation}
S(k) = \frac{1}{N}\langle \sum_{ij} \mathrm{e}^{-ik(r_j - r_i)} \rangle, 
\end{equation}
with $|{\bf k}|\equiv k$ a crystal vector, and characterizes diagonal order. We use up to $N$=188 particles and about 750 sites to minimize finite-size effects and defect concentrations of up to four percent.\\

\begin{figure}
\centerline{\includegraphics[width=0.8\columnwidth]{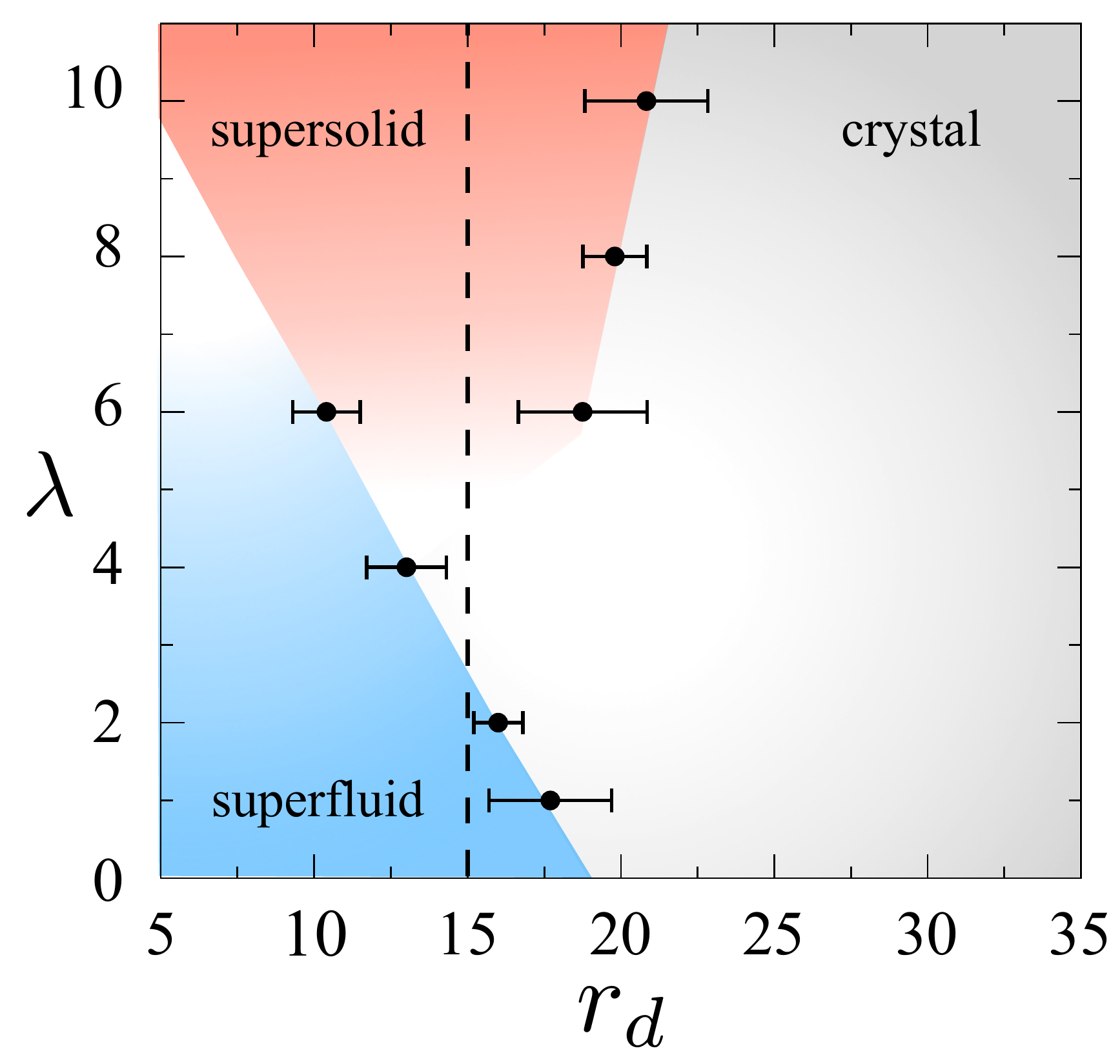}}
\caption{(Colour online) Phase diagram $r_d$ vs. $\lambda$ for Eq.~(1).
$f_s$ and $S^{max}$ for $r_d=15$ (dashed line) are reported in Fig.\ref{fig:phasediagram1}, see text.
Error bar on $r_d$ has been estimated using a fine size scaling analysis.}
\label{fig:phasediagram}
\end{figure}

Quantum phases of Eq.~\eqref{hamiltonian} with $\lambda=0$ have been investigated for the case of bosonic polar molecules in Refs.~\cite{Buechler, Astra2007,Filinov,Moroni2014}. The phase diagram is characterised by a melting quantum phase transition from a triangular crystal phase [with $S^{max}\neq0$ and $f_s=0$] to a homogeneous superfluid [with $S^{max}=0$ and $f_s\neq0$] by decreasing the interaction strength, defined as
\begin{equation}
\label{RD}
r_d=\frac{\Gamma_0 m}{a \hbar^2},
\end{equation}
below a critical value $r_d^c= (18\pm3)$. Here we are interested in investigating the phase diagram as a function of 
$r_d$ and of the depth $\lambda$ of an additional optical lattice.\\

Figure~\ref{fig:sff1} (upper panel) displays the ground state limit of $f_s$ vs. $\lambda$ for an interaction strength $r_d=7.5 < r_d^c$, corresponding to a superfluid for $\lambda=0$. In the figure, we consider both the case of a lattice commensurate with the dipole density at filling $n = 1/4$ (black circles) and the case of non-commensurate filling with a small density of interstitial defects (equaling 0.02 and 0.04 for the red empty and full squares, respectively). 

For $n=1/4$ (black circles), the figure displays a sudden drop of $f_s$, suggesting a quantum phase transition from a homogeneous superfluid ($f_s =1$) to an insulating phase ($f_s=0$)  induced by the increase of the strength of the lattice potential, occurring at the critical value $\lambda\approx 8$. This picture is corroborated by a sudden increase of the static structure factor at the same value $\lambda \simeq 8$ [Fig.~\ref{fig:sff1} (lower panel)], showing that the insulating phase is in fact a crystal. This transition is thus driven by the suppression of quantum kinetic energy with increasing $\lambda$~\cite{Bloch2008}. As we discuss below, for physical realizations with, e.g.,  polar molecules KRb or Er$_2$, the value of $\lambda\simeq 8$ corresponds  to a very small value of the lattice depth, where the two-dimensional band structure is not formed. Thus, this quantum phase transition can be regarded to happen in the continuum, as a two-dimensional analog of the so-called pinning quantum phase transition that has been predicted in one dimension~\cite{Dalmonte,Buchler2003, Giamarchi}. For {\it short-range} interactions, the one-dimensional analog has been observed experimentally in 1D~\cite{ExpNaegerl,Buchler2003} while the theree-dimensional case has been studied via quantum Monte-Carlo simulations in Ref.~\cite{Pilati}.\\

The scenario of the superfluid-crystal quantum phase transition described above changes drastically when defects are introduced: in this case $f_s$ remains finite but  not unitary, as expected for a {\it non-homogeneous} superfluid \cite{leggett}. This is shown in Fig.~\ref{fig:sff1} (upper panel)] in the parameter regime $8 \lesssim \lambda \lesssim 12$,  where the superfluid density increases with the concentration of interstitial defects (see, in particular, the cases $\lambda=10$ and 12). 

\begin{figure}[b]
\centerline{\includegraphics[width=0.8\columnwidth]{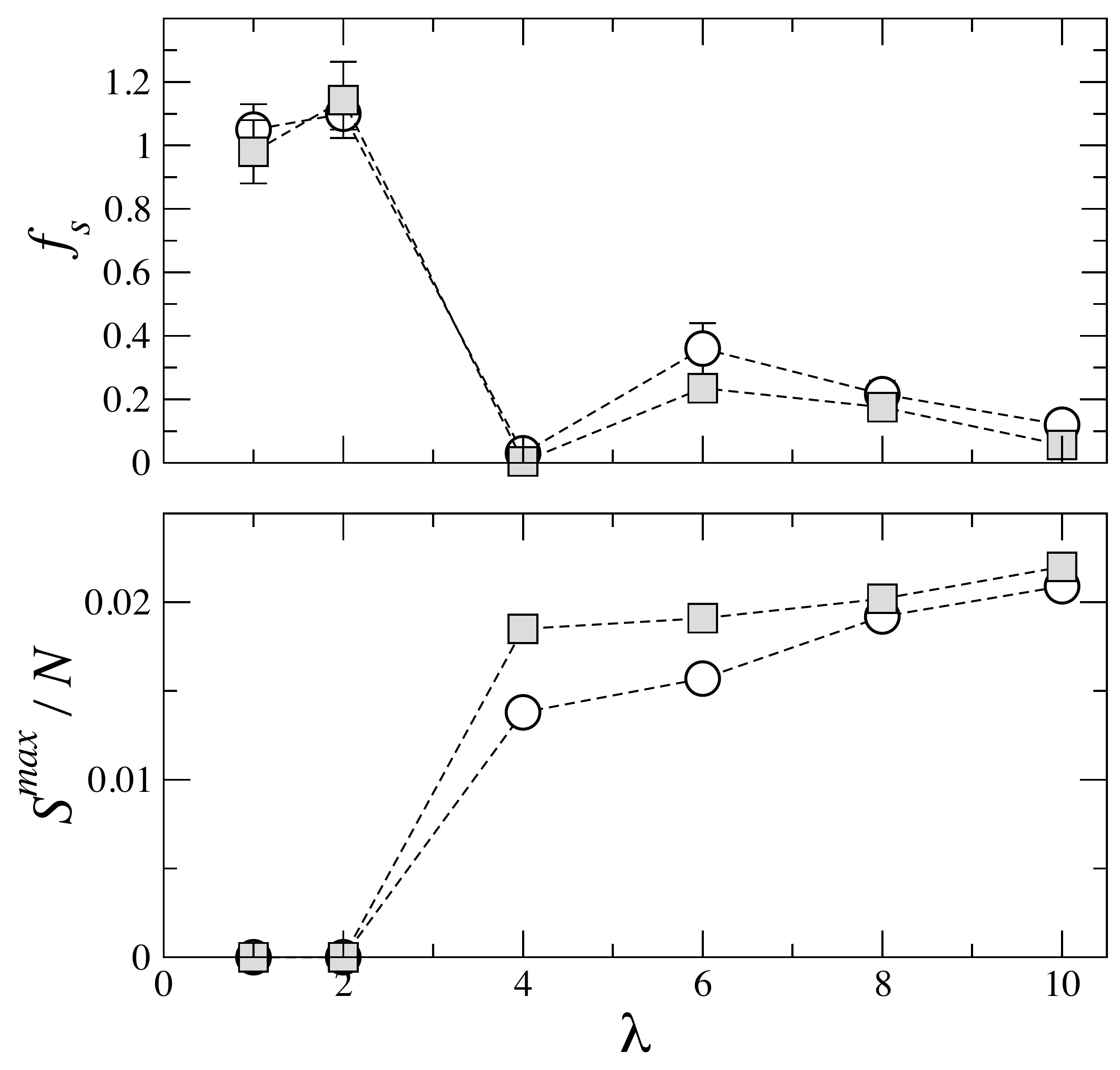}}
\caption{$f_s$ and $S^{\max}$ vs. $\lambda$ for $N=26$ (circle) and $N=52$ (square) considering $r_d=15$ (see dashed line in Fig.~\ref{fig:phasediagram}).}
\label{fig:phasediagram1}
\end{figure}

Interestingly, we find that in the whole parameter range $\lambda\gtrsim8$ the order parameter $S^{max}$ is finite and maximal at the crystal wave-vector $\textbf{k}=(4\pi/3a_0,0)$ corresponding to the {\it commensurate} dipolar crystal with $n=1/4$ (as in the case described above), implying a diagonal crystalline order with a periodicity that is {\it different} from that one of the underlying lattice potential \cite{Note1}. This is in contrast, e.g., to the case with short-range interactions~\cite{Pilati}, where the periodicity of the (Mott) insulating phase is trivial, in that it coincides with the one of the underlying lattice. 

Here, the coexistence of a finite $S^{max}$ and $f_s>0$ in the parameter regime with $\lambda > 8$ demonstrates the realisation of a supersolid state of matter with coexisting diagonal order and superfluidity.\\

Figure~\ref{fig:phasediagram} shows the complete zero-temperature phase diagram as a function of $r_d$ and $\lambda$, again keeping the defect density fixed at 0.02. As described above, we find regions of superfluid, crystal and supersolid behaviour. In particular, for $\lambda=0$ we re-obtain the phase diagram for a bosonic dipolar gas discussed above~\cite{Buechler}, characterised by the quantum melting transition of the triangular crystal into a homogeneous superfluid at $r_d^c \simeq 18$. We find that increasing $\lambda$ from $0$ initially has the simple effect to shift the quantum melting transition to smaller values of $r_d < r_d^c$, consistent with the example of Fig.~\ref{fig:sff1} (black empty dots).  However, the nature of the transition changes dramatically above $\lambda>5$: a defect-induced supersolid phase intervenes between the superfluid and the crystal. For $\lambda \gtrsim 6$ all phases can be observed by simply tuning $r_d$.

An interesting example of this latter situation is shown in Fig.~\ref{fig:phasediagram1}, which presents results for $r_d=15$, corresponding to a cut in the phase diagram of Fig.~\ref{fig:phasediagram} along the indicated dashed line. Figure~\ref{fig:phasediagram1} shows that, while $S^{\rm max}$ grows monotonically with increasing $\lambda \gtrsim 2$, the superfluid density $f_s$ displays an interesting {\it re-entrant behavior}: it is $f_s\simeq 1$ for $\lambda<2$, then drops to zero for   $\lambda \simeq 4$, and then becomes again finite with $f_s \simeq 0.2$ - $0.4$ for $\lambda \gtrsim 6$; this latter behavior corresponds to a supersolid.
We notice that in this parameter region with $r_d<r_d^c$ the crystalline structure is purely imposed by the presence of the lattice potential, which pins the strongly interacting superfluid, however, crucially without opening a gap. This is similar to the lattice supersolid discussed in Refs.~\cite{Pollet2010,Boninsegni2005}, however it occurs for shallow lattice depths, where the band structure is not formed (see below).

For $r_d>r_d^c$ the triangular crystalline structure is present also for $\lambda=0$. As the crystal is essentially classical (however, see~\cite{HEXTRANS}),  we expect that the results on the tunability of defect interactions derived in the Sec.III above should provide directly insights into defect dynamics in this parameter regime. Indeed, in the PIMC quantum calculations we find that the effect of a sufficiently deep lattice in this crystalline case (e.g.,  $\lambda \gtrsim 8$) is to originate a finite superfluid density coexisting with crystalline order, when a finite density of defects is present. This defect-induced supersolidity is only possible for effective repulsive interactions between the defects, as would be predicted by the classical results given above. 

Finally, for sufficiently large $\lambda \gtrsim 11$ the superfluid fraction vanishes altogether, and the groundstate evolves into an insulating lattice-type crystal. This is similar to the observed frozen dynamics in the classical regime. \\

\begin{figure}[t]
\centerline{\includegraphics[width=0.99\columnwidth]{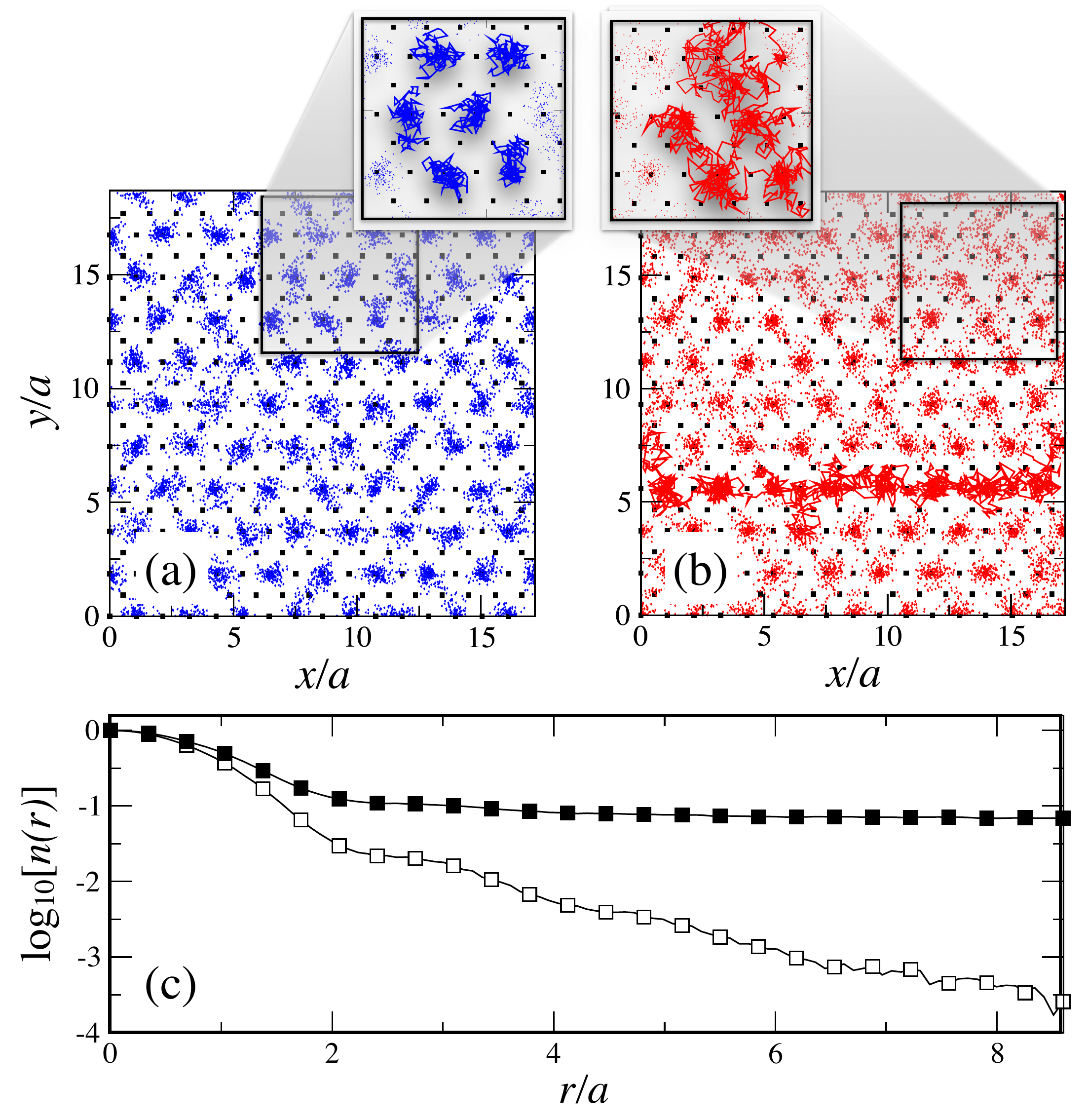}}
\caption{ (Colour online)  Quantum Monte-Carlo snapshots with 
$N$=80 (a) and $N$=88 (b) particles with $r_d\approx7.5$
(320 sites, square points) and $\lambda$=10. In the presence of interstitials (right panel) delocalized paths emerge (red thick line) and result in a finite $f_s$. 
(c) Spherical averaged one-body density matrix (with same set of parameter of panels a and b)
for $n$=1/4 (crystal, open symbol) and for a supersolid phase on lattice (full symbol). 
The error bars lie within point size.}
\label{fig:conf}
\end{figure}

In order to visualise the difference between the solid and supersolid phases, Fig.~\ref{fig:conf} shows snapshots of the projection of world-lines onto the xy-plane taken from the PICM simulations, obtained by tracing over imaginary time, at  $n=1/4$ in the absence of defects [Fig.~\ref{fig:conf}(a)] and with two interstitials [Fig.~\ref{fig:conf}(b)] for $r_d=15$ and $\lambda=10$. As explained in literature~\cite{Ceperley}, these projections (that are for illustration purposes only) are the closest representation of the square of the wave-function for the many-body system that can be obtained in a simulation, where overlapping paths imply exchanges among the bosonic particles and superfluidity. The figure shows that in the absence of defects, paths remain localised around the local minima of the lattice potential $U(x,y)$. However, in the supersolid phase localised paths coexist with paths that are delocalized throughout the system, representing cyclic exchanges (permutations) among bosons [see Fig.~\ref{fig:conf}(b), red thick line]. This residual exchange mechanism is consistent with defect-induced supersolidity as originally proposed by Andreev and Lifshitz~\cite{ALC2} and only recently demonstrated via exact theoretical techniques for bosons with cluster-forming interactions~\cite{CINTI}. 
Our analysis shows that defect-induced supersolidity can be originated {\it in the continuum} also for non-cluster forming liquids, using periodic external potentials. Apart of the qualitative difference, as explained below, this should provide quantitative advantages in the experimental realisation of the supersolid phase, as it could results in, e.g, higher temperatures than possible in the tight-binding regime. 
\\

Particle delocalization is also reflected in the (\textit{quasi}-) condensate fraction, which is easily accessible in experiments~\cite{Bloch2008}, defined as the asymptotic (i.e., $r\rightarrow \infty$) behaviour of the  angle averaged one-body density matrix 
\begin{equation}
\label{nr}
n(r) = \frac{1}{2\pi V}\int d\Omega \int d{\bf r}\,n({\bf r},{\bf r}^{\prime}),
\end{equation}
 with
 \begin{equation}
n({\bf r},{\bf r}^{\prime})=\langle\hat{\psi}({\bf r}\,)\hat{\psi}^\dagger({\bf r}^{\prime}) \rangle,
\end{equation}
and  $\hat{\psi}({\bf r})$  [$\hat{\psi}^\dagger({\bf r})$]  the particle annihilation [creation] operators at position 
${\bf r}$ \cite{Pethick}. In the presence of long-range off-diagonal order associated to a finite {\it condensate} fraction,  $n({\bf r},{\bf r}^{\prime})$ factorizes at large separation $|{\bf r}-{\bf r}^{\prime}|$ as 
 \begin{equation}
\langle\hat{\psi}({\bf r}\,)\hat{\psi}^\dagger({\bf r}^{\prime}) \rangle \rightarrow \phi({\bf r})\phi({\bf r}^{\prime}),
\end{equation}
with $\phi({\bf r})$ the condensate wave-function. Employing the same set of parameters of Fig.~\ref{fig:conf}(b),
Fig.~\ref{fig:conf}(c) shows that a constant value for the supersolid phase with finite defect 
concentration is here realised (full squares), corresponding to a finite (\textit{quasi}-)condensate fraction at $T\simeq 0$. The latter disappears in the case of commensurate filling $n=1/4$ [parameters as in Fig.~\ref{fig:conf}(a)], where $n({\bf r},{\bf r}^{\prime})$ decays exponentially with distance [empty square in panel (c)], as expected for an insulating crystalline phase.\\

\begin{figure}[b]
\centerline{\includegraphics[width=0.9\columnwidth]{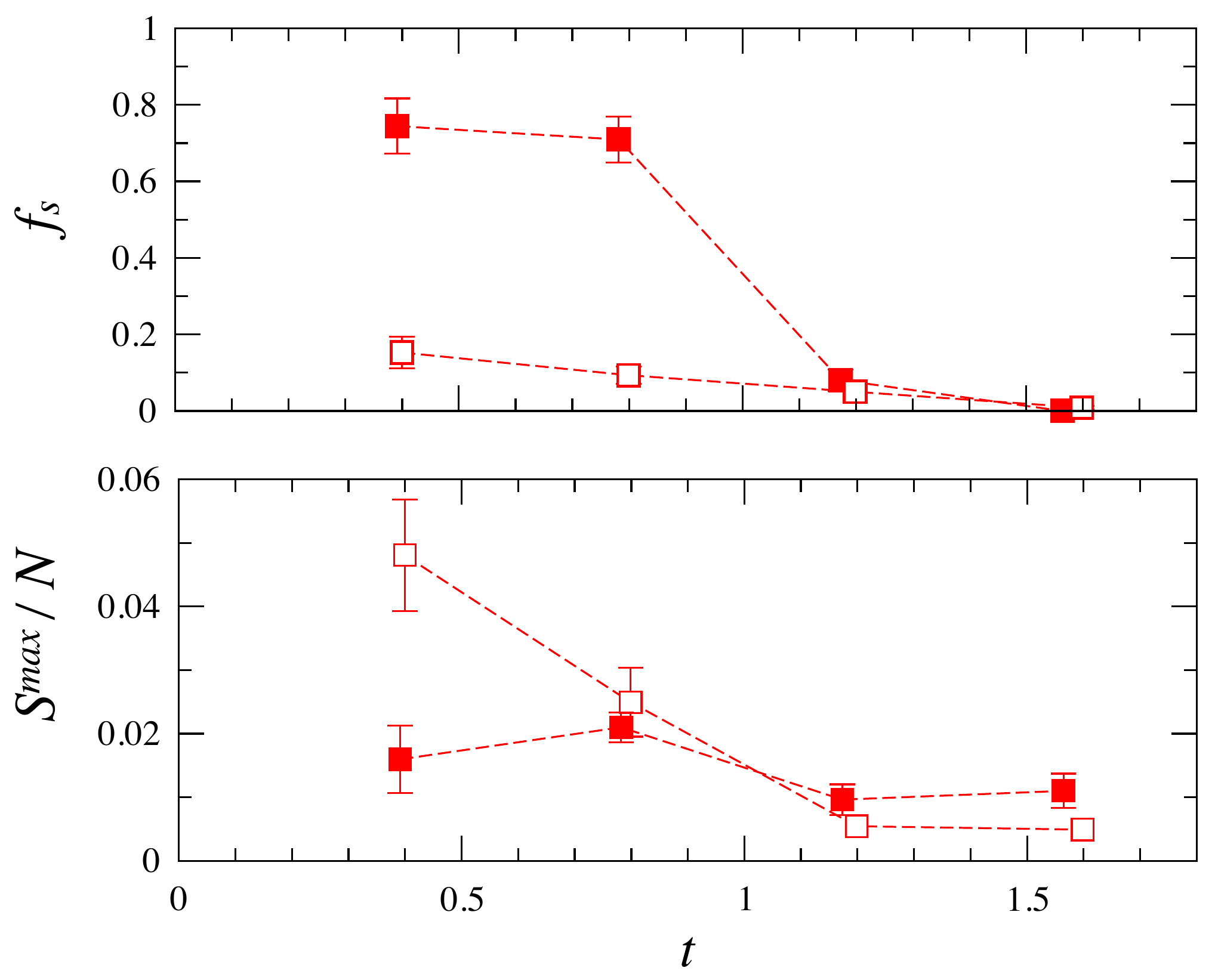}}
\caption{(Colour online) Superfluid fraction (top panel) and Structure factor (lower panel) 
of the reduced temperature $t=k_B T/(\hbar^2n/m)$ 
with $\lambda$=10 and  $N_d=4$ (full square, interstitial density 0.02), and $N_d=8$ (open square, interstitial density 0.04).}
\label{fig:sff2}
\end{figure}

\section{Finite temperature and possible physical realizations}

The supersolid phase described here is considerably resilient towards finite temperature effects. An example of this is shown in Fig.~\ref{fig:sff2}, where we present results for both $f_s$ and $S^{max}$ as a function of the rescaled temperature  $t \equiv k_B T/(\hbar^2 n/m)$, expressed in units of the quantum kinetic energy at the mean inter particle distance $\hbar^2 n/m$. The figure shows that supersolid behaviour survives up to temperatures of the order of   $t \equiv k_B T/(\hbar^2 n/m)\lesssim$1. The latter is consistent with a BKT transition for $f_s$ with a comparatively large transition temperature $T_{\rm BKT}$ (see below).\\

\subsection{Polar molecules:  $KRb$}
As an example of experimental realisation of the phase above, here we consider first a gas of bosonic $^{39}$K$^{85}$Rb molecules with dipole moment $d\simeq 0.5$ Debye trapped on a lattice with spacing $a_0=400$ nm. The lattice recoil energy is $E_R / h= h/(8 m a_0^2)\simeq 2.49$ kHz (in frequency units), while the unit of energy $\Gamma = \Gamma_0/a_0^3$ reads $\Gamma / h \simeq 0.59$ kHz. By re-expressing the value $\lambda=10 \Gamma$ in convenient $E_R$ units, the numbers above suggest that the supersolid phase described above would be realised for a weak lattice potential with a depth of just $\gtrsim2.3 E_R$. This corresponds to a situation where the band structure of the two-dimensional lattice has not formed, which is consistent with our claim that this quantum phase transition occurs in the continuum. 
The quantum kinetic energy at the average particle density $E_{\rm kin}/h= \hbar n / (2 \pi m_{\rm KRb})$ in the absence of the optical lattice reads $E_{\rm kin}/h \simeq 130$ Hz. . 
Thus, from the results of Fig.~\ref{fig:sff2} we obtain that the supersolid phase would survive up to temperatures of the order of $T_c\simeq 6$ nK or larger. 

\subsection{Magnetic quantum gases: $Er_2$}
Magnetic quantum gases of $Er$ have been recently trapped in lattices with a small spacing $a_0=266$nm~\cite{Ferlaino2015-2}. For $Er_{2}$ molecules with magnetic dipole moment of $Er_{2} = 14 \mu_B$ (with $\mu_B$ the Bohr magneton) this implies $\Gamma / h \simeq 0.14$kHz, which is comparable to the case of polar molecules. 
Extrapolation from our numerical results (see Fig.~\ref{fig:phasediagram}) implies that a lattice depth of order of $\lambda\simeq 10\Gamma$ is necessary to induce supersolidity. 

Here, the recoil energy is $E_R\simeq 2.1$kHz, and thus the condition on the lattice depth $\lambda$ for inducing a supersolid behaviour reads $\lambda \simeq 10 \Gamma\simeq 0.7E_R$, which is well in the continuum limit. The quantum kinetic energy reads $E_{\rm kin}/h= \hbar n / (2 \pi m_{Er_2})\simeq 105$Hz, for a low density $n=1/(4 a_0^2)$.  This corresponds to a critical temperature for observing supersolidity of the order of $T_c\gtrsim 5$nK, which is within experimental reach. We note that for the case of strongly dipolar magnetic gases, reaching densities of order unity in the lattice is possible. This in principle would allow for the observation of some of the effects described above, such as the superfluid-insulator quantum phase transition for commensurate lattice fillings, at considerably higher temperatures. 

\section{Conclusions and Outlook}
The results presented here may help to settle the long-standing question of the role of defect interactions for the formation of the supersolid phase \cite{DEFECTS_FATE,ANDERSON} and enable the experimental realization of a supersolid with quantum dipolar gases in combination with tunable optical lattices. We note that while here we have focused on specific examples with fixed defect densities, defect-induced supersolidity is expected to appear for a large range of densities for a careful choice of interaction parameters, e.g., similar to the case of Ref.~\cite{CINTI}. This should make the observation of defect-induced supersolidity possible in experiments where particle density is tunable within a few percent only. The tunability of defect interactions opens also interesting prospects for the observation of other phases, such as solitons and breathers in the classical and quantum regimes~\cite{BlatterPRB2014,Dalmonte}. \\

{\textit Acknowledgements} - We thank M. Troyer, P. Zoller, T. Macr\`i and F. Mezzacapo for fruitful discussions. WL acknowledges support by the Austrian Science Fund (FWF): P 25454-N27. GP is supported by the ERC-St Grant ColdSIM (No. 307688), EOARD, RySQ, UdS via IdEX and ANR via BLUESHIELD.

\bibliographystyle{prsty}

\end{document}